\documentclass[aps,twocolumn,showpacs]{revtex4}
\usepackage{amssymb}
\usepackage{amsmath}
\usepackage{graphicx}

\begin{document}

\title{Comparison of Supernovae datasets Constraints on Dark Energy}

\author{Zhang Cheng-wu}
\author{Xu Li-xin}
\author{Chang Bao-rong}
\author{Liu Hong-ya}

\affiliation{School of Physics and Optoelectronic Technology, Dalian
University of Technology, Dalian,116024, P.R. China}

\pacs{98.80.Es, 98.80.-k, 95.36.+x}
\begin{abstract}
Cosmological measurements suggest that our universe contains a
dark energy component. In order to study the dark energy
evolution, we constrain a parameterized dark energy equation of
state $w(z)=w_0 + w_1 \frac{z}{1+z}$ using the recent
observational datasets: 157 Gold type Ia supernovae and the newly
released 182 Gold type Ia supernovae by maximum likelihood method.
It is found that the best fit $w(z)$ crosses $-1$ in the past and
the present best fit value of $w(0)<-1$ obtained from 157 Gold
type Ia supernovae. The crossing of $-1$ is not realized and
$w_0=-1$ is not ruled out in $1\sigma$ confidence level for the
182 Gold type Ia supernovae. We also find that the range of
parameter $w_0$ is wide even in $1\sigma$ confidence level and the
best fit $w(z)$ is sensitive to the prior of $\Omega_m$.
\end{abstract}
\maketitle

The universe is currently accelerating revealed by recent
observations of type Ia supernovae (SNe Ia)\cite{Riess98,Perlmutter}
and the acceleration has been attributed to a mysterious component
dubbed dark energy. Although a cosmological constant $\Lambda$ is
the simplest candidate for dark energy and appears to explain
current observations satisfactorily\cite{WMAP03,WMAP06}, it suffers
from fine tuning and coincidence problems\cite{Peebles03}. These
difficulties have lead to a large variety of proposed models with
time-depended dark energy, such as quintessence\cite{quintessence},
phantom\cite{phantom}, quintom\cite{quintom},
K-essence\cite{K-essence}, tachyonic matter\cite{tachyon} and so on.
In all the above models, the evolution history of the universe and
the nature of dark energy strongly depend on the models, such as the
dynamics of the extra energy component and the gravity theory used
to solve the cosmological problem. However, there is another
approach to study dark energy properties directly by observations in
a model independent manner. In this method, the equation of state of
dark energy $w(z)=p(z)/\rho(z)$ could be tested by cosmological
observations to understand  the gravitational properties of dark
energy and its evolution. The aim of this paper is to compare the
157 Gold SNe Ia (SN157)\cite{Riess04} with the latest 182 Gold SNe
Ia (SN182)\cite{Riess06} in constraining dark energy with a
parameterized equation of
state\cite{Eos1-a,Linder03,Choudhury,Nesseris}
\begin{equation}
w(z)=w_0 + w_1 \frac{z}{1+z}\label{wz}
\end{equation}
where $w_0$ and $w_1$ are constant, $z$ is redshift. Since the
spatially flat universe $\Omega_m+\Omega_{DE}=1$ agrees well with
observations\cite{WMAP03,WMAP06}, in this paper, we just consider
the flat Friedmann-Robertson-Walker cosmology.

In a flat universe with Eq.(\ref{wz}), the Friedmann equation can
be expressed as
\begin{equation}\label{hz}
H^2 (z)=H_0^2 [ \Omega_{m} (1+z)^3
+\Omega_{DE}(1+z)^{3(1+w_0+w_1)}e^{\frac{-3w_1 z}{(1+z)}}]
\end{equation}
Then the knowledge of $\Omega_{m}$ and $H(z)$ is sufficient to
determine $w(z)$.

Supernovae observations provide the apparent magnitude $m(z)$ at
peak brightness after implementing correction for galactic
extinction, K-correction and light curve width-luminosity
correction. $m(z)$ is related to the luminosity distance $d_L(z)$
through

\begin{eqnarray}
m_{th}(z)&=&5\log_{10} (D_L (z))+ M + 5
\log_{10}(\frac{H_0^{-1}}{Mpc}) + 25 \nonumber\\
&=&5\log_{10} (D_L (z))+ M + 42.38 - 5 \log_{10}h \label{mdl}
\end{eqnarray}
where $H_0=h \, km\cdot s^{-1}\cdot Mpc^{-1}$, parameter $M$ is the
absolute magnitude which is assumed to be constant and $D_L(z)$ is
defined as
\begin{eqnarray}
D_L(z)&=&H_{0}d_L(z)/c\nonumber\\&=&(1+z)\int_{0}^{z}\frac{H_{0}dz^{'}}{cH(z^{'})}
\end{eqnarray}Supernovae data points are given in terms of the
distance modulus
\begin{equation}
\mu_{obs}(z_i)\equiv m_{obs}(z_i) - M
\end{equation} and theoretical model parameters
are determined by minimizing the quantity
\begin{equation}
\chi^2= \sum_{i=1}^N \frac{(\mu_{obs}(z_i) -
\mu_{th}(z_i))^2}{\sigma_{(obs; i)}^2}  \label{chi2}
\end{equation}where $\sigma_{(obs; i)}^2$ are the errors due to flux uncertainties, intrinsic
dispersion of SNe Ia absolute magnitude and peculiar velocity
dispersion respectively. These errors are assumed to be gaussian
and uncorrelated. The theoretical distance modulus is defined as
\begin{eqnarray}
\mu_{th}(z_i)&\equiv& m_{th}(z_i) - M \nonumber\\ &=&5 \log_{10}
(D_L (z)) +42.38 - 5 \log_{10}h \label{mth}
\end{eqnarray}
 and $\mu_{obs}$ is given by supernovae dataset.

\begin{figure}[!htbp]
  \includegraphics[width=0.45\textwidth]{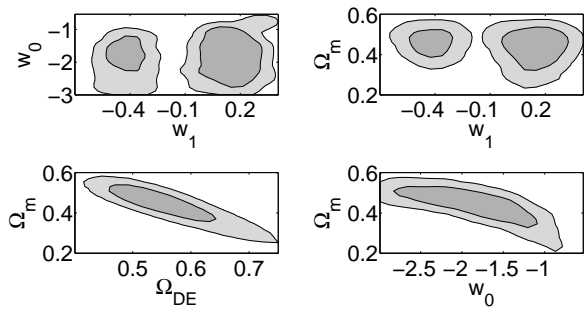}\\
  \caption{The contours show 2-D marginalized $1\sigma$ and
$2\sigma$ confidence limits in the ($w_1$, $w_0$), ($w_{1}$,
$\Omega_{m}$), ($\Omega_{DE}$, $\Omega_{m}$), ($w_{0}$,
$\Omega_{m}$) plane from SN157.}\label{157}
\end{figure}

\begin{figure}[!htbp]
  \includegraphics[width=0.4\textwidth]{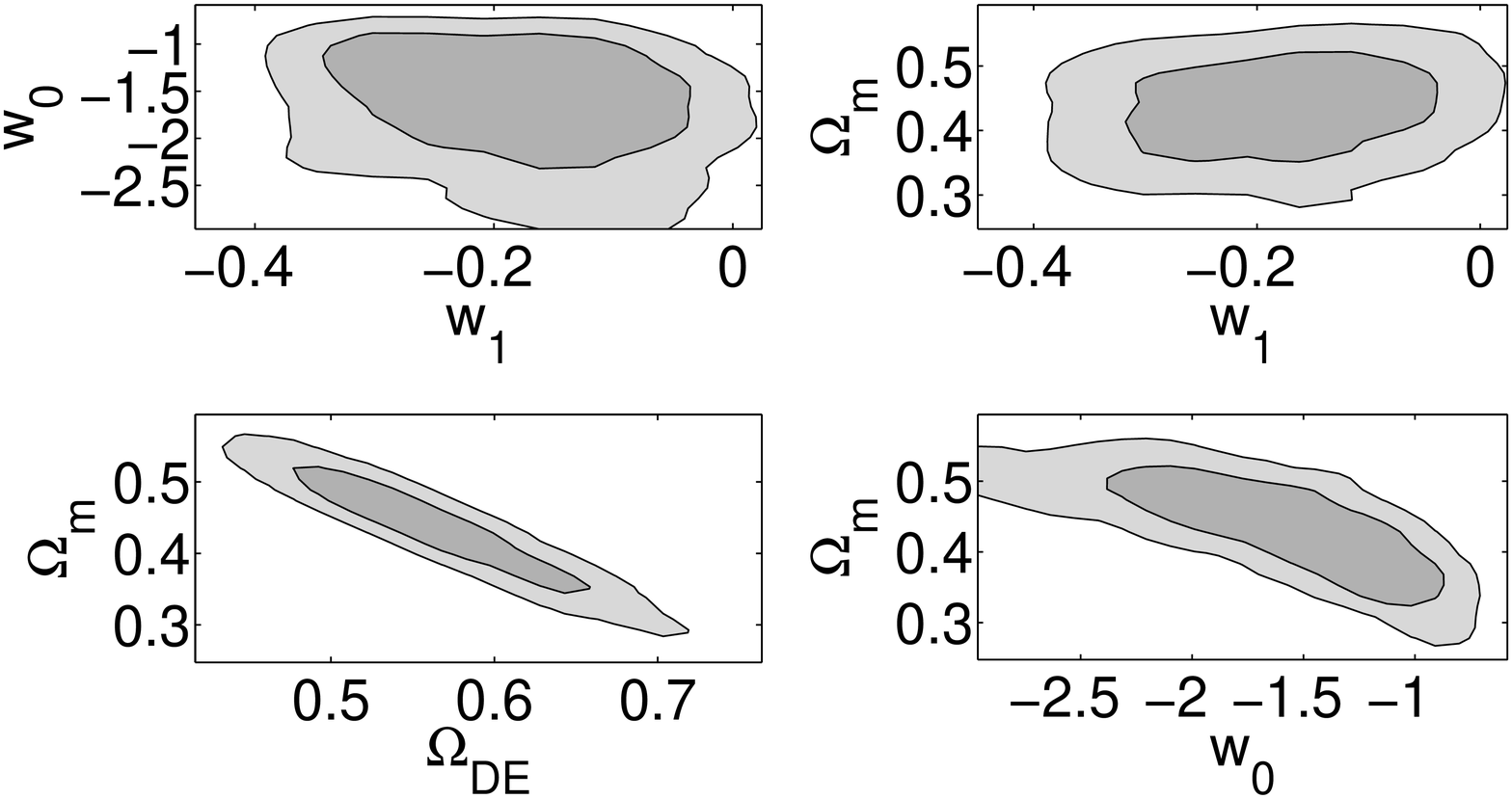}\\
  \caption{The contours show 2-D marginalized $1\sigma$ and
$2\sigma$ confidence limits in the ($w_1$, $w_0$), ($w_{1}$,
$\Omega_{m}$), ($\Omega_{DE}$, $\Omega_{m}$), ($w_{0}$,
$\Omega_{m}$) plane from SN182.}\label{182}
\end{figure}

\begin{figure}[!htbp]
  \includegraphics[width=0.45\textwidth]{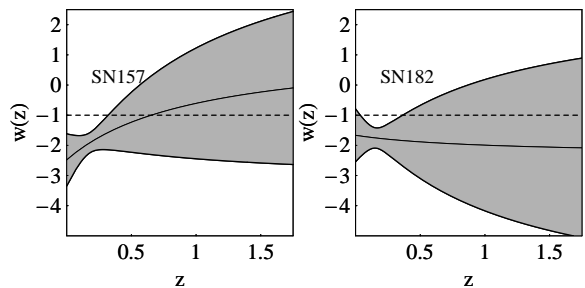}\\
  \caption{The best fits of $w(z)$ with $1\sigma$ errors (shaded region).
   $\Omega_m$ is set to the best values obtained from SN157 or SN182.}\label{w-z}
\end{figure}

\begin{figure}[!htbp]
  \includegraphics[width=0.45\textwidth]{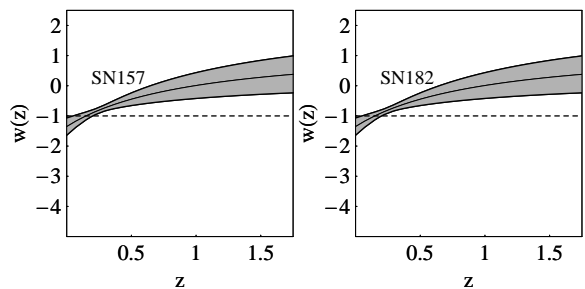}\\
  \caption{The best fits of $w(z)$ with $1\sigma$ errors (shaded region) with a prior $\Omega_m=0.27$.}\label{w-z2}
\end{figure}

\begin{table}\label{tab1}
\noindent TABLE 1. The minimum $\chi^2$ and mean fit values  from
SN157 and SN182.
\begin{center}
\begin{tabular}{c c c c c}

\hline
data &  $\chi^2_{min}$  &  $w_0$& $w_1$ &$\Omega_{m}$\\
\tableline
  SN157 & 89.947 & $-1.856 \pm 0.519$ & $-0.015 \pm 0.269$ & $0.437 \pm 0.073$ \\
  SN182 & 81.523 & $-1.637 \pm 0.496$ & $-0.180 \pm 0.081$ & $0.435 \pm 0.058$ \\
\hline
\end{tabular}
\end{center}
\end{table}

Our calculations to limit parameters are based on the Markov Chain
Monte Carlo program CosmoMC\cite{Lewis02}. We modified the code to
make equation of state $w$ be redshift-depended instead of
constant $-1$ and add parameter $w_0$ and $w_1$ as slow parameters
to the cosmological parameters space. Using the maximum likelihood
method with SN157 and SN182, we obtained confidence level of
equation of state and matter density. The contour plots are show
in Fig.\ref{157} and Fig.\ref{182}, the mean fit values, $1\sigma$
constraints on parameters $w_0$, $w_1$ and $\Omega_m$ with
standard deviations are list in Table 1. In Fig.\ref{157} we show
results for the parameterized equation of state for the SN157
dataset. We find the best fit value of $(w_1,w_0,\Omega_m)$ to be
$(0.335,-2.373,0.485)$ and see that the $(w_1,w_0)$ panel has two
island separately as well as the $(w_1,\Omega_m)$ panel which
means that the SN157 dataset can not give a fine constraint on
$w_1$. Fig.\ref{182} shows the results calculated by the SN182
dataset. The best fit value of $(w_1,w_0,\Omega_m)$ is
$(-0.244,-1.728,0.462)$.

Fig.\ref{w-z} shows $1\sigma$ errors of the best fit $w(z)$
calculated using the covariance matrix\cite{AlamU04}. The best fit
$w(z)$ constrained by SN157 can cross $-1$ in the evolution and in
$1\sigma$ confidence level, the LCDM parameter ($w_0=-1,w_1=0$) do
not coincide with the best fit value of $w(z)$ at present. But the
best fit dynamical $w(z)$ obtained from SN182 does not cross $-1$
for the whole redshift range $0<z<1.75$, $w_0=-1$ is in $1\sigma$
confidence contour and near the boundary. During our calculations,
we also find that the best fit $w(z)$ and its errors  are
sensitive to the value of $\Omega_m$. When we set $\Omega_m=0.27$
as a prior instead of the best fit value mentioned before, both of
the best fit $w(z)$ can cross $-1$ and its $1\sigma$ error region
shrink sharply as showed in Fig.\ref{w-z2}.

In summary, we have provided a comparative analysis  on
constraining parameters $w_1, w_0$ , $\Omega_m$ and the properties
of dark energy with two cosmological observations: SN157 and
SN182. It shows that the SN157 dataset can not give a fine
constraint on $w_1$ and the best fit $w(z)$ can cross $-1$ in the
evolution, but the SN182 dataset gives contrary result. We also
find that the range of parameter $w_0$ is wide even in $1\sigma$
confidence level and the best fit $w(z)$ is sensitive to the prior
of $\Omega_m$.  In the future, we can combine other cosmological
observations, such as Cosmic Microwave Background
anisotropies\cite{WMAP06}, the SDSS baryon acoustic
peak\cite{SDSS}, the cluster baryon fraction (CBF)\cite{Allen04}
or the linear perturbations growth rate at $z=0.15$ obtained from
the 2dF galaxy redshift survey (PGR)\cite{Verde02} to constrain
the matter density and the evolution of dark energy. We hope that
the best fit values of cosmological parameters and the confidence
level could be enhanced which will help us to understand the
nature of dark energy.

\textbf{Acknowledgements} We acknowledge the using of
CAMB\cite{camb} and CosmoMC\cite{cosmomc} for the numerical
calculations. Chengwu Zhang thanks the helpful discussion with
Prof. Antony Lewis\cite{cosmocoffee} and suggestions from Dr.
Haitao Cui. This work was supported by NSF (10573003), NBRP
(2003CB716300) of P.R. China. The research of Lixin Xu was also
supported in part by NSF (10647110) and DUT 893321.


\begin{thebibliography}{99}

\bibitem{Riess98}
Riess A~G et al 1998 {\em Astron. J.\/} {\bf 116} 1009
  astro-ph/9805201

\bibitem{Perlmutter}
Perlmutter S et al 1999 {\em
  Astrophys. J.\/} {\bf 517} 565 astro-ph/9812133

\bibitem{WMAP03}
Spergel D~N et al 2003 {\em Astrophy. J. Suppl.\/} {\bf
  148} 175 astro-ph/0302209

\bibitem{WMAP06}
Spergel D~N et al 2006 {\em arXiv\/}
  Astro-ph/0603449

\bibitem{Peebles03}
Peebles P~J~E and Ratra B 2003 {\em Rev. Mod. Phys.\/} {\bf 75}
559
  astro-ph/0207347

\bibitem{quintessence}
Zlatev I, Wang L~M and Steinhardt P~J 1999 {\em Phys. Rev.
Lett.\/} {\bf 82}
  896

\bibitem{phantom}
Caldwell R~R, Kamionkowski M and Weinberg N~N 2003 {\em Phys. Rev.
Lett.\/}
  {\bf 91} 071301

\bibitem{quintom}
Feng B, Wang X~L and Zhang X~M 2005 {\em Phys. Lett. B\/} {\bf
607} 35

\bibitem{K-essence}
Armendariz-Picon C, Damour T and Mukhanov V 1999 {\em Phys. Lett.
B\/} {\bf
  458} 209

\bibitem{tachyon}
Padmanabhan T 2002 {\em Phys. Rev. D\/} {\bf 66} 021301

\bibitem{Riess04}
Riess A~G et al 2004 {\em Astrophys. J.\/} {\bf 607} 665
astro-ph/0402512

\bibitem{Riess06}
Riess A~G et al 2006 {\em
  arXiv\/} Astro-ph/0611572

\bibitem{Eos1-a}
Chevallier M and Polarski D 2001 {\em Int. J. Mod. Phys. D\/} {\bf
10} 213
  gr-qc/0009008

\bibitem{Linder03}
Linder E~V 2003 {\em Phys. Rev. Lett.\/} {\bf 90} 091301
astro-ph/0208512

\bibitem{Choudhury}
Choudhury T~R  and Padmanabhan T 2005  {\em Astron.Astrophys.} {\bf
429} 807 astro-ph/0311622


\bibitem{Nesseris}
Nesseris S and Perivolaropoulos L 2005 {\em Phys. Rev. D\/} {\bf 72}
123519 astro-ph/0511040


\bibitem{Lewis02}
Lewis A and Bridle S 2002 {\em Phys. Rev. D\/} {\bf 66} 103511

\bibitem{AlamU04}
Alam U et al 2004 {\em ArXiv\/}
  Astro-ph/0406672

\bibitem{SDSS}
Eisenstein D~J~ et al 2005 {\em Astrophys. J.\/} {\bf 633} 560

\bibitem{Allen04}
Allen S~W~ et al 2004{\em Mon. Not. R. Astron. Soc.} {\bf 353} 457
astro-ph/0405340

\bibitem{Verde02}
Verde L et al 2002 {\em Mon. Not. R. Astron. Soc.} {\bf 335} 432
astro-ph/0112161

\bibitem{camb}
  Http://camb.info

\bibitem{cosmomc}
  Http://cosmologist.info/cosmomc

\bibitem{cosmocoffee}
  Http://cosmocoffee.info

\end{thebibliography}

\end{document}